\documentstyle[epsfig]{l-aa}

\newcounter{saveeqn}
\newcounter{cl}

\newcommand{\alpheqn}[1]{\setcounter{saveeqn}{\value{equation}}%
\setcounter{cl}{\value{saveeqn}}%
\stepcounter{saveeqn} \refstepcounter{cl}%
\setcounter{equation}{0}%
\renewcommand{\theequation}{\mbox{\arabic{saveeqn}\alph{equation}}}
\label{#1}}

\newcommand{\reseteqn}{\setcounter{equation}{\value{saveeqn}}%
\renewcommand{\theequation}{\arabic{equation}}}

\newcommand{\beq}[1]{\begin{equation} \label{#1}}
\newcommand{\eeq}{\end{equation}}

\newcommand{\FWX}{{\rm FWHM_{X-ray}}}
\newcommand{\FW}[1]{{\rm FWHM_{#1Hz}}}

\newcommand{\FWHMm}{{\rm FWHM_m}}
\newcommand{\FWHMn}{{\rm FWHM_n}}

\newcommand{\Flux}[1]{{\rm S_{#1Hz}}}

\newcommand{\Sm}{S_m}
\newcommand{\Sn}{S_n}

\newcommand{\Imm}{I_m}
\newcommand{\In}{I_n}

\newcommand{\epsm}{\epsilon_m}
\newcommand{\epsn}{\epsilon_n}
\newcommand{\qmn}{q_{mn}}
\newcommand{\alphaSmn}{\alpha_{S, mn}}
\newcommand{\alphaImn}{\alpha_{I, mn}}
\newcommand{\alphaemn}{\alpha_{\epsilon, mn}}
\newcommand{\rmn}{r_{\rm mn}}

\newcommand{\BCMB}{B_{\rm CMB}}

\newcommand{\muG}{\mu{\rm G}}

\begin{document}

\thesaurus{02.01.1; 02.04.2; 11.03.4 NGC 1656; 11.09.3; 13.18.1}
\title{The large-scale structure of the diffuse radio halo of the
Coma cluster at 1.4 GHz}
\author{B.M.~Deiss\inst{1}
\and
W.~Reich\inst{2}
\and
H.~Lesch\inst{3}
\and
R.~Wielebinski\inst{2}}
\institute{Institut f\"ur Theoretische Physik, J.W.~Goethe-Universit\"at,
  Robert-Mayer-Stra{\ss}e 10, \\D-60054 Frankfurt am Main,
  Germany
\and
  Max-Planck-Insitut f\"ur Radioastronomie, Auf dem H\"ugel 69,
  D-53121 Bonn, Germany
\and
  Institut f\"ur Astronomie und Astrophysik der Universit\"at M\"unchen,
  Scheinerstra{\ss}e 1,\\ D-81679 M\"unchen, Germany}
\offprints {B.M.~Deiss}
\date{Received date;accepted date}
\maketitle

\begin{abstract}
We present new measurements of the diffuse radio emission from the
Coma cluster of galaxies at 1.4 GHz using the Effelsberg 100-m-telescope.
Even at that high frequency,
the halo source Coma C has an extent down to noise of $\sim 80\arcmin$
corresponding to 3 Mpc (${\rm H_0 = 50 kms^{-1}Mpc^{-1}}$) in Coma.
The radio map reveals clear similarities with images of the
extended X-ray halo of the Coma cluster. However, the radio halo
appears to be displaced from the X-ray halo by $\sim\, 3-4$ arcmin.
After subtracting the contributions from point sources we obtained
an integrated diffuse flux density of $\Flux{1.4G} = 640 \pm 35$ mJy
from Coma C.

We derive relations between the various observationally determined
spectral indices and the spectral index of the synchrotron emissivity,
which allow one to achieve a rough estimate concerning
the consistency of the presently available data at different
frequencies and to place constraints on the emissivity index distribution.
In the halo's core region, the inferred emissivity index between 0.3 GHz and
1.4 GHz appears to be in the range 0.4 - 0.75 implying that
there must be some very effective mechanism for particle acceleration
operating in the intracluster medium. We discuss implications
of our measurements and of the spectral index information on current
theories for radio halo formation.
We stress the importance of having
more measurements of Coma C at frequencies above 1.4 GHz, in order to
be able to derive constraints on the physics of the
formation process of the radio halo.

\end{abstract}

\section{Introduction} \label{introduct}

Diffuse cluster-wide radio emission not associated with individual
galaxies defines a separate class of extragalactic radio sources:
the diffuse radio halos of galaxy clusters. Radio halo sources
are observed in the richest  and most X-ray luminous clusters of
galaxies. However, one of their most enigmatic properties appears
to be their obvious rarity. At present, only a few clusters are
definitely known to have an extended radio halo: e.g., A754
(Waldthausen 1980), A2255 (Harris et al. 1980),
A2256 (Bridle \& Fomalont 1976),
A2319 (Harris \& Miley 1978). The best studied example, however, is the
Coma cluster (A1656) with its extended, diffuse halo source Coma C
(e.g.,Hanisch 1980; Hanisch \& Erickson 1980; Waldthausen 1980; Cordey 1985;
Schlickeiser et al. 1987; Henning 1989; Venturi et al.
1990; Kim et al. 1990; Giovannini et al. 1993).

At lower frequencies
diffuse radio emission from Coma C has been
observed up to an angular distance of $\sim 35\arcmin$
from the cluster center [e.g., Henning (1989) at 30.9 MHz], which
corresponds to a radius of 1.5 Mpc in Coma (z = 0.0235; Sarazin
et al. 1982).(We adopt $H_0 = 50 {\rm km\,s^{-1}}$ for the
Hubble constant throughout the paper; i.e., $1\arcmin \cor
40$ kpc). This implies the existence of a cluster-wide presence
of relativistic electrons as well as of a magnetic field.

Kim et al. (1990) published
a detailed image of Coma C at 1.4 GHz using a synthesis aperture
telescope (Dominion Radio Astrophysical Observatory, DRAO).
Applying a Gaussian fit, they inferred a $\FW{1.4G} $ of
the radio halo of $18\farcm7 \times 13\farcm7$ which is significantly
smaller than the value of the FWHM obtained at lower frequencies,
e.g., at 326 MHz (Venturi et al. 1990).
{}From these observations Giovannini et al. (1993) concluded that the
spectral index of the diffuse radio emission strongly steepens with
increasing radius.
The DRAO radio map of Kim et al.
(1990) reveals a halo diameter down to noise level of $< 25\arcmin$.
However, emission from larger structures could be attenuated due to
the synthesis aperture technique. This has implications for the
determination of the spectral index distribution (Giovannini et al. 1993)
as well as  of the integrated flux at 1.4 GHz.
In order to obtain informations on the large scale characteristics
of the halo at high frequencies, we observed Coma C at 1.4 GHz
with the Effelsberg single-dish 100-m-telesope.

The exact shape of the diffuse radio emission spectrum of Coma C is yet
unclear.
Schlickeiser et al. (1987) claimed that the integrated flux density spectrum
strongly steepens at high frequencies ($\nu > 1$ GHz).
We discuss the consistency of their measurements to our observations,
and we discuss some implications of our new measurements on current
models for radio halo formation which have been proposed in the literature
(e.g. Hanisch 1982, Tribble 1993).

\section{Radio continuum observations}

\begin{figure}
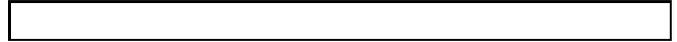

 \picplace{0.5cm}
  \caption[]{\label{fig:contour_orig} 1.4 GHz map of the Coma cluster
   from Effelsberg 100-m telesope observations. The rms noise is
   7 mJy/beam. Contours are 10 mJy/beam apart (dashed contour: 0.0 mJy/beam).
   The HPBW (9\farcm35) is indicated in the lower left-hand corner
   }
\end{figure}

We have made radio continuum observations at 1.4 GHz with the Effelsberg
100-m telescope in April 1993. At this frequency the  telescope has an
angular resolution (HPBW) of $9\farcm35$. A field of $3\degr \times
3\degr$ centered on the Coma cluster has been mapped twice in
orthogonal directions. We used a two channel receiver with cooled HEMT
amplifiers. A bandwidth of 20 MHz was centered on 1.4 GHz. The data have
been processed using standard procedures for continuum mapping observations
with the Effelsberg 100-m telescope (e.g. Reich et al., 1990). The data
from both channels have been averaged and the two coverages have been
combined using the method described by Emerson and Gr\"ave (1988).
The final map is limited by confusion and has an r.m.s.-noise of about
7 mJy (or 14.4 mK Tb). We displayed our result in the form of a contour
plot in Fig. \ref{fig:contour_orig}. The map shows numerous compact radio
sources, most of them associated with Coma cluster galaxies, superimposed
on a weak large scale diffuse emission component.

\begin{figure}
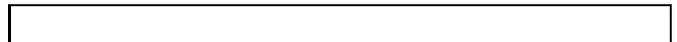

 \picplace{0.5cm}
  \caption[]{\label{fig:contour_halo} 1.4 GHz map of the Coma cluster
   as in Fig. \ref{fig:contour_orig} but with compact sources subtracted.
   Contours are 10 mJy/beam apart (dashed contour:  0.0 mJy/beam).
   The HPBW (9\farcm35) is indicated in the lower left-hand corner
   }
\end{figure}

In order to separate the diffuse extended emission from the contribution
of individual sources, we have used the master list of radio sources from
the Coma cluster as compiled by Kim (1994). The data for 298 sources
are from various observations made with synthesis telescopes and therefore
include sources as weak as a few mJy not directly accessible by us due
to our larger beam width and resulting higher confusion limit. We have
used the spectral fits by Kim et al. (1994) to calculate the flux density
of all sources at 1.4 GHz and subracted these contributions assuming
a Gaussian source shape at the listed positions. We have in addition
subtracted a few sources fitted by a Gaussian at the edge areas of our
field which are outside the region where Kim (1994) has listed radio
sources. The result of this procedure is shown in Fig. \ref{fig:contour_halo}
where a weak diffuse radio component is left which is centered on the
Coma cluster.

The east-west angular extent of the diffuse radio source is more than
80\arcmin.
The bridge-like extension of the diffuse radio emission to the south-west
is related to the galaxy group associated with the bright galaxy NGC 4839.
In order to obtain the integrated flux density of the diffuse radio emission
at 1.4 GHz from Coma C, we integrated the diffuse flux
(Fig. \ref{fig:contour_halo}) over a circular area of radius $40\arcmin$
centered
at $\alpha = 12^h57^m10\fs7$, $\delta = 28\degr12\arcmin16\arcsec$ (1950), but
where we subtracted contributions from the region around NGC 4839 and from the
narrow halo extension to the north. The latter might be related to the
radio source 5C4.109.


\section{Results}

\subsection{Morphological structure of the radio halo}\label{morph}

At 1.4 GHz, the diffuse radio halo Coma C reveals a butterfly-like
shape with  an angular extent down to noise of more than
$80\arcmin$ along the east-west direction and of about $45\arcmin$
in north-south direction, corresponding to 3.2 Mpc $\times$ 1.8 Mpc
respectively. The 1.4 GHz aperture synthesis map
of Coma C recently obtained by Kim et al. (1990) shows an
angular size of the radio halo down to noise
of only 22\arcmin, due to the low sensitivity to
the very large scale radio emission.

An even more striking feature is the close resemblance of the
spatial structure of the halo source at 1.4 GHz and the smooth,
extended X-ray halo of Coma (Briel et al. 1992; White et al. 1993).
The X-ray emission is thermal bremsstrahlung originating from the
hot ($T \sim 10^8$ K) intracluster gas (Sarazin 1988). Both the
X-ray halo and the 1.4 GHz radio halo are elongated approximately
east-west. Both the ROSAT PSPC X-ray image (White et al. 1993) and
our 1.4 GHz halo map (Fig. \ref{fig:contour_halo}) reveal a narrow
extension towards the galaxy group associated with NGC 4839 south-west of the
Coma cluster. Our radio map actually shows that this extended structure
represents a narrow, low-brightness bridge of diffuse radio emission
connecting the radio source Coma C with the peripheral extended complex
of radio emission, Coma A.
This bridge was also detected at 326 MHz by Venturi et al.
(1990).

Fig. \ref{fig:halo_profil} shows the azimuthally averaged surface brightness
distribution of the 1.4 GHz map of the halo centered at
$\alpha = 12^h57^m10\fs7$, $\delta = 28\degr12\arcmin16\arcsec$ (1950).
For comparison, we plotted the model function (normalized to 200 mJy)
given by Briel et al. (1992) which fits the azimuthally
averaged surface brightness distribution of the X-ray halo observed
with ROSAT.
As a model function, Briel et al. (1992) adopted a modified isothermal
King profil (King 1966) $I(\Theta)/I_0 = (1+(\Theta/a)^2)^{-(3\beta - 1/2)}$,
where
$\Theta$ denotes the projected radius, $a$ is the core radius, $\beta$ is
the density slope parameter, and $I_0$ is the central surface brightness.
The radial profil of the X-ray surface brightness distribution is fitted
best adopting a $\beta = 0.75 \pm 0.03$ and a core radius
$a = 10\farcm5 \pm 0\farcm6$ (Briel et al. 1992). The galaxy distribution
of the cluster is more centrally concentrated having a core radius
of 6-8 arcmin (Sarazin 1988, and references therein). The FWHM of the
X-ray halo is $\FWX = 14\farcm6$, while one infers a $\FW{1.4G} = 15\farcm2$
from
the azimuthally averaged, deconvolved brightness distribution at 1.4 GHz.
This shows that the scale size of the diffuse radio halo source at 1.4 GHz
is similar to that of the X-ray halo. However, the radio emission declines
 more rapidly in the outer regions of the cluster.
\begin{figure}
\epsfig{bbllx=3cm, bblly=5cm, bburx=19cm, bbury=21cm,file=MS4661Fig3.eps,
width=9cm}
  \caption[]{\label{fig:halo_profil} Azimuthally averaged surface brightness
   distribution of the diffuse radio halo of the Coma cluster at 1.4 GHz
(dots).
   The measurement points are connected by straight solid lines, in order to
have
   a better impression about the radial profil. Dashed curve: Best fitting
   King model of the radial profil of the X-ray halo (Briel at al. 1992)
   }
\end{figure}

As has already been noted by Kim et al. (1990), the positions of the two
sources are significantly displaced. On the largest scale, we find the radio
source to be about 3-4 arcmin west of the X-ray source. The location of the
peak surface brightness in our map (Fig. \ref{fig:contour_halo}) seems to
be even more displaced. This must not be taken too seriously,
considering i) the low angular resolution of the observation, and ii)
the sensitivity of the result of the removal procedure
concerning the proper positioning of the strong radio sources
5C4.85 and 5C4.81 in the center region of the cluster.
Inspecting the 2.7 GHz map of the diffuse radio halo obtained by
Schlickeiser et al. (1987), one finds that also at this high frequency
the halo source reveals
a positional offset of 2-3 armin to the west relativ to the X-ray source.
{}From the existence of the positional offset between radio halo
and X-ray halo, Kim et al. (1990) concluded that the relativistic
particles in the radio halo may not be directly responsible for
the heating of the X-ray emitting gas. Since, if the relativistic
particles were the main heating source of the hot gas, the two
sources should be precisely coextensive. Nevertheless, the close
morphological correspondence between thermal X-ray
bremsstrahlung and non-thermal synchrotron radiation
suggests that the existence of the diffuse radio emission is based on the
physical condition of the hot thermal gas.

\subsection{Integrated diffuse flux from Coma C}

Performing the procedure described in Sect. 2 we obtained an
integrated diffuse flux density from Coma C  of
$\Flux{1.4G} = 640 \pm 35$ mJy. Kim et al. (1990) derived an integrated
diffuse flux of $530 \pm 50$ mJy at 1380 MHz using the DRAO Synthesis
Telescope and the NRAO Very Large Array.  Our single-dish observation,
however, indicates that the radio halo at 1.4 GHz is much more
extended than it is suggested by the DRAO image,
which accounts for the increased integrated
flux at that frequency. Fig. \ref{fig:spectrum} shows the integrated
diffuse flux from Coma C at various frequencies. The data set used
(see Table \ref{table:spectrum_data}) comprises our measurement
together with
data used by Schlickeiser et al. (1987) and by Giovannini et al. (1993).
Our measurement seems to fit
rather neatly a power law extrapolation from lower frequencies.
The data points in the frequency range
$\lid 1.4$ GHz may be fitted by a power law with index
$\alpha = 1.16 \pm 0.03$. If we take all data into account a power-law
fit gives $\alpha = 1.36 \pm 0.03$.
\begin{table}
  \caption[]{\label{table:spectrum_data}
   Integrated flux densities from Coma C. References: (1) Henning
   1989; (2) Hanisch \& Erickson 1980; (3) Cordey 1985; (4) Venturi
   et al. 1990; (5) Kim et al. 1990; (6) Hanisch 1980;
   (7) Giovannini et al. 1993;
   (8) present paper; (9) Schlickeiser et al. 1987; (10) Waldthausen
   et al. 1980
   }
  \begin{flushleft}
  \begin{tabular}{lllllllllllllllllll}
  \hline
  & Frequency (MHz) & Flux (Jy) & Error (Jy) & References &\\
  \hline
  &30.9     & 49    & 10    & 1  &\\
  &43       & 51    & 13    & 2  &\\
  &73.8     & 17    & 12    & 2  &\\
  &151      & 7.2   & 0.8   & 3  &\\
  &326      & 3.18  & 0.03  & 4  &\\
  &408      & 2.0   & 0.2   & 5  &\\
  &430      & 2.55  & 0.28  & 6  &\\
  &608.5    & 1.2   & 0.3   & 7  &\\
  &1380     & 0.53  & 0.05  & 5  &\\
  &1400     & 0.64  & 0.035 & 8  &\\
  &2700     & 0.070 & 0.020 & 9  &\\
  &4850     &$<$ 0.052 & ...   & 10 &\\
  \hline
  \end{tabular}
  \end{flushleft}
\end{table}
\begin{figure}
 \epsfig{bbllx=3cm, bblly=8cm, bburx=19cm, bbury=24cm,file=MS4661Fig4.eps,
width=9cm}
  \caption[]{\label{fig:spectrum}
   Integrated flux density spectrum of the diffuse radio halo Coma C.
   Data are from Table
   \ref{table:spectrum_data}.
   }
\end{figure}

\subsection{Scale-size-vs.-frequency relation}\label{section_size_freq}

Our value for the FWHM of the halo at 1.4 GHz (15\farcm2) is in agreement with
the scale size obtained by Kim et. (1990) ($18\farcm7 \times
13\farcm7$), who fitted the brightness distribution with a
two-dimensional Gaussian. At 326 MHz, Venturi et al. (1990)
inferred a $\FW{326M} = 28\arcmin \times 20\arcmin$. Henning (1989)
derived the scale size of the radio halo at an even lower
frequency (30.9 MHz). She found $\FW{30.9M} = 31\arcmin \pm 5\arcmin
\times 15\arcmin \pm 5\arcmin$.
In order to obtain a rough estimate of the halo's FWHM at 2.7 GHz, we
azimuthally averaged the 2.7 GHz map given by Schlickeiser et
al. (1987) and inferred a FWHM from that radial distribution.
This yields a $\FW{2.7G} \sim
12\farcm6 \pm 1\farcm3$ which may be regarded as an upper limit, since we
did not deconvolve the original map. In any case, this is
significantly smaller than the scale size of the halo at 1.4 GHz.
Hence, there appear to be strong observational evidences that
the scale size of the radio halo is a monotonically decreasing
function of frequency.
This supports the suggestion made by Giovannini et al. (1993)
that in the external regions of the cluster the diffuse radio halo
exhibits a steeper spectrum than in the core region. Comparing
surface brightness measurements at 326 MHz and 1.4 GHz, performed
with synthesis aperture telescopes, Giovannini et al.
(1993) derived a spectral index distribution which shows an almost
constant index of $\sim 0.8$ within a cluster radius of $\sim 8\arcmin$
and a strong increase of the index up to values higher than 1.8
outside this central "plateau". However, our single-dish observations
indicate that at larger spatial scales there is much more diffuse radio
power at 1.4 GHz than it is suggested by the synthesis aperture measurements.
Thus,
the increase of the spectral index may be weaker than claimed
by Giovannini et al. (1993). A smaller spectral index implies
a weaker net energy loss of the relativistic electrons in the peripheral
region of the cluster.

\section{Discussion}
\subsection{Consistency of data and the spectral index distribution}

The integrated diffuse flux measurements listed in Table
\ref{table:spectrum_data}
may be subjected to considerable systematic errors: for instance,
an improper subtraction of contributions from
point sources, a too small integration volume, or an erroneous
intensity offset of the diffuse emission.
Due to the very low number of data points, these (unknown)
systematic errors may dominate any statistical analysis
of a theoretical halo model. In that respect, it still seems to
be an open question whether the strong steepening of the integrated-flux
spectrum of Coma C above 1 GHz,
claimed by Schlickeiser et al. (1987), is real. Our observations
indicate that, at least up to 1.4 GHz, the integrated spectrum shows no
tendency to steepen. If the 2.7 GHz value obtained by Schlickeiser et al.
(1987) is real
this implies a rather distinct cutoff of the spectrum between
1.4 GHz and 2.7 GHz. However, such a sharp spectral break can hardly be
reproduced
by any theoretical halo model, since, even in the case that the electron
energy distribution tends to zero at some characteristic energy, the
decline of the synchrotron emission spectrum is at most exponential.
Hence, one may suspect that at frequencies $> 1.4$ GHz the
diffuse radio halo is much more luminous  than it is suggested by the
measurements by Schlickeiser et al. (1987) and Waldthausen (1980).

Using the combined data of the integrated flux densities,
the surface brightness distributions and the scale-size-vs.-frequency
relation
one may achieve, under some simplifying assumptions, a rough estimate
concerning the consistency of the measurements at different frequencies.
In addition, the considerations described in the following allow
to place some constraints on the spectral index
distribution of the synchrotron emission coefficient.

The FWHM at a given frequency is usually inferred from a Gaussian fit
to the spatial profil of the surface brightness distribution. This appears to
be
reasonable, since the shape of the surface brightness distribution is more
or less Gaussian. This implies that the spatial profil of the emission
coefficient $\epsilon_{\nu}(\vec{r})$ may be assumed to be
Gaussian-shaped, too. Taking advantage of that allows one to derive a relation
between the spectral index distribution of
$\epsilon_{\nu}$ and other spectral indices which can be determined by
observations.
For the sake of simplicity, we consider a  spherically symmetric halo.
At a given frequency, say $\nu = m$, the integrated diffuse radio
flux density spectrum $\Sm$ and the surface brightness distribution
$\Imm(b)$ at a projected radius $b$ are given by
\beq{def:S}
  \Sm = ({\rm const}) \,\int\limits_0^{\infty} r^2 \epsm (r)\,{\rm d}r
\eeq
and
\beq{def:I}
  \Imm(b) = ({\rm const})\,\int\limits_0^{\infty} \epsm(r^2=z^2+b^2)\,{\rm d}z.
\eeq
In Eqs. (\ref{def:S}) and (\ref{def:I}) we implicitely assumed a quasi-locally
averaged,
isotropic emission coefficient which depends only on the absolute value
 of the cluster radius $r$. This is a reasonable simplification, since
the cluster magnetic field which generally introduces a directional
dependence appears to be tangled on rather small scales.
Feretti et al. (1995) give scale sizes for the magnetic field reversals
of less than 1 kpc; this value is required to explain the dispersion of the
rotation measure of the radio source NGC 4869.

For two different frequencies, say $\nu_1 = m$ and $\nu_2 = n$, and
on the assumption that the radial distributions
of $\epsm(r)$ and $\epsn(r)$ may be reasonably well fitted by Gaussians,
the following relations hold:
\beq{S_ratio}
  \frac{\Sm}{\Sn} =
\frac{\epsm(0)}{\epsn(0)}\,\left(\frac{\FWHMm}{\FWHMn}\right)^3
\eeq
and
\beq{eps_ratio}
 \frac{\Imm(b)}{\In(b)} = \left.\frac{\epsm(r)}{\epsn(r)}\right|_{r=b}
                             \,\frac{\FWHMn}{\FWHMm}.
\eeq

As it is usually done, we characterize the
ratios $\Sm/\Sn$, $\Imm(b)/\In(b)$ and $\epsm(r)/\epsn(r)$
by spectral indices $\alphaSmn$, $\alphaImn(b)$ and $\alphaemn(r)$,
respectively.
In addition, we define a corresponding index $\qmn$ for the size-vs.-frequency
relation
through the ratios
\beq{q_def}
  \frac{\FWHMm}{\FWHMn} = \left(\frac{m}{n}\right)^{-\qmn}.
\eeq

Employing Eqs. (\ref{S_ratio}) and (\ref{eps_ratio}) we can now relate
the spectral index $\alphaemn(r)$ of the emission coefficient to the
observationally
given indices $\alphaSmn$, $\alphaImn(b)$ and $\qmn$.
We obtain
\beq{alphS}
   \alphaemn(0) = \alphaSmn - 3\,\qmn,
\eeq
\beq{alpheps}
   \left.\alphaemn(r)\right|_{(r=b)} = \alphaImn(b) - \qmn.
\eeq
In the latter equation, the notation $(r=b)$ means that
the value of the cluster radius $r$, at which we would like
to know the index $\alphaemn(r)$, is set equal to the
value of the projected radius $b$ at which the index
$\alphaImn(b)$ is measured.

Equations (\ref{alphS}) and (\ref{alpheps})  imply the relation
\beq{alphSI}
   \alphaSmn = \alphaImn(0) + 2\,\qmn
\eeq
which provides
a check on the consistency of the observational data. It
relates to each other the measurements of the integrated flux, of the
central surface brightness, and of the scale size of the halo at two
different frequencies. Of course, considering the employed simplifications
such a relation will never be exactly fullfilled. Nevertheless, it can
give a hint to whether there are considerable systematic errors,
e.g., in the
measurement of the integrated flux densities, as discussed above.

Since we assumed  that the surface brightness distributions have
Gaussian-shaped profils,
the theoretical spectral-index distribution $\alphaImn(b)$ is
necessarily a quadratic function given by
\alpheqn{alphI}
\beq{}
  \alphaImn(b) = \alphaImn(0) + \frac{b^2}{\rmn^2},
\eeq
where  $\rmn$ is defined through
\beq{}
  \rmn = (4{\rm ln 2})^{-1/2}\FWHMm\left[{\rm ln}\frac{n}{m}\right]^{1/2}
         \left[\left(\frac{\FWHMm}{\FWHMn}\right)^2 - 1 \right]^{-1/2}
\eeq
\reseteqn
At the characteristic radius $\rmn$ the spectral index  is greater
than at the cluster center by one.
The comparison of the theoretical profil (\ref{alphI}) and the observed
spectral-index distribution may serve as an additional check on
the consistency of the data; and it
gives a hint on the degree of reliability of the Gaussian fits.

As a first example, we consider the
observations at $m = 326$ MHz (Venturi et al. 1990) and $n = 1.38$ GHz
(Kim et al. 1990)
(see Table \ref{table:spectrum_data} and Sect. \ref{section_size_freq}) .
Inserting $\FW{326M} \simeq 24\arcmin$ and $\FW{1.38G} \simeq 16\arcmin$
in definition (\ref{q_def})
we obtain $\qmn = 0.28$. The spectral index of the integrated flux
is $\alphaSmn = 1.23$.
Then, for reasons of consistency,
the central value of the spectral index of the
surface brightness $\alphaImn(0)$ should be 0.67 according to relation
(\ref{alphSI}).
This value is slightly smaller but still in good
agreement with the value measured by Giovannini et al. (1993) (0.6 - 0.8;
their Fig. 4). Hence, according to (\ref{alpheps}), the spectral index of the
emission coefficient at the cluster's center is $\alphaemn(0) \approx 0.4 -
0.5$.
{}From Eq. (\ref{alphI}b) we infer a characteristic radius of
the spectral-index distribution of $\rmn = 15\farcm6$; i.e., the
spectral index (\ref{alphI}a) increases from its central value of 0.67 to
a value of 0.93 at a projected radius of $b = 8\arcmin$
(the central "plateau" according to Giovannini et al. 1993),
and it reaches 1.8 at
$b = 16\farcm6$ . This shows that
the theoretical spectral-index distribution
fits the main features of the observed one;
hence, the
assumption of Gaussian surface brightness profils does not introduce artificial
inconsistencies.
Thus, within the frame of our simplifying assumptions, the observational data
at 326 MHz and 1.38 GHz appear to be consistent.

As a second example, we consider the Effelsberg observations at
$m = 1.4$ GHz (present work) and $n = 2.7$ GHz (Schlickeiser et al. 1987).
{}From the measurements of the integrated flux densities one infers
a value for the spectral index of $\alphaSmn = 3.37$.
Taking $\FW{1.4G} = 15\farcm2$ and $\FW{2.7G} \simeq 12\arcmin$, we obtain
$\rmn = 9\farcm5$ and
$\qmn = 0.36$; the latter implies that, according to (\ref{alphSI}),
 the spectral index of the surface brightness
at $b = 0$ should have a value of $\alphaImn(0) = 2.65$.
The peak surface brightness of the 2.7 GHz map is $\sim 11$ mJy/beam
(beamwidth $4\farcm3$).
The peak surface
brighness of our 1.4 GHz map (Fig. \ref{fig:contour_halo}) is $\sim 130$ mJy,
where the beamwidth is $9\farcm35$; this would reduce roughly by a factor
of $(4\farcm3/9\farcm35)^2 = 0.21$ if one used a $4\farcm3$ beam.
Hence, one derives a value of the surface-brightness index of
$\alphaImn(0) \sim 1.40$ which is considerably smaller than that
expected from relation (\ref{alphSI}).
Our 1.4 GHz observations appear to be in accord with the measurements
made by Kim et al. (1990), as has been discussed above.
Hence, the disagreement between the indices strongly indicates that
the 2.7 GHz data are inconsistent, in the sense that the value of
the integrated
diffuse flux given by Schlickeiser et al. (1987) is, most probably, too low.
Since the observed surface brightness at 2.7 GHz declines rapidly in the
peripheral region
of the halo, an extension of the integration area, which might have been
too small, would lead to an increase
of the total flux of only a few percent. Another source of error, to which
the integrated-flux measurement at 2.7 GHz is very sensitive due to the
low surface brightness, is the determination of
the correct intensity offset of the diffuse emission. For instance,
one would yield an additional integrated diffuse flux of $\sim 50$ mJy
if one lowered the intensity offset by only 1 mJy/beam; however, the
resulting spectral indices were still inconsistent in that case.
The data would be roughly consistent, if
the surface brightness at 2.7 GHz were about 2.5 mJy/beam higher
than the values given in the 2.7 GHz map;
the integrated
diffuse flux density would then presumably amount to
$S_{\rm 2.7 GHz} \sim 200$ mJy and the values
of the spectral indices would be $\alphaSmn \approx 1.8$ and $\alphaImn(0)
\approx 1.1$, while the FWHM would increase only slightly.
In that case, $\alphaemn(0)$ were about 0.75 at the cluster's center.

Since the scale size of the halo appears to be a decreasing function
of frequency, one expects, for reasons of consistency, an increase
of the spectral index of the integrated diffuse flux density.
However, the extremly sharp spectral break
above 1.4 GHz, suggested by the presently available integrated-flux
data, seems to be unrealistic. Nevertheless, above 1.4 GHz
the spectral index of the emission coefficient seems to increase
even in the core region of the cluster.

\subsection{Implications on theories of halo formation}

In this section, we discuss some implications on theories of radio halo
formation
following from the considerations of the previous section and from
the observed large extent of the radio halo at 1.4 GHz and its clear similarity
with the X-ray halo.

The dominant energy loss processes of the relativistic electrons
in the intracluster medium are synchrotron emission and  inverse
Compton  scattering off
the cosmic microwave background radiation (CMB).

The "lifetime" of
an (isotropic) ensemble of relativistic electrons radiating at
1.4 GHz is given by (Pacholczyk 1970)
\beq{def:tau_loss}
  \tau_{\rm loss} = 1.3\,10^8\, \left(\frac{B}{\mu{\rm G}}\right)^{1/2}
                   \left(1 + \frac{2}{3}\frac{B^2}{\BCMB^2}\right)^{-1}
                   \,{\rm yr},
\eeq
where the "monochromatic approximation" is used, and where
$\BCMB = 3.18(1+z)^2\mu$G
denotes the magnitude of the magnetic field equivalent to the CMB.
The magnetic field strength in the intracluster
medium of the Coma cluster is still a matter of debate.
Kim et al. (1990) derived a magnetic field strength of
$B = 1.7 \pm 0.9 \mu$G using  excess Faraday rotation measure (RM)
of polarized emission from background radio sources.
Recently, however,
Feretti et al. (1995) inferred a much stronger
magnetic field of $\sim 6 \pm 1\,h_{50}^{1/2} \mu$G
from polarization data of the radio galaxy NGC 4869 located in
the central region of the cluster.
Hence, the electrons' lifetime seems to be at most $10^8$ years.

This short lifetime and the observed large extent of the 1.4-GHz halo
augment the well-known
diffusion speed problem of the primary-electron model suggested
by Jaffe (1977) and Rephaeli (1979): In order to reach the edge
of the halo, "primary electrons" which are presumed to be
ejected from central radio galaxies must propagate at a speed of at
least $v_{\rm prop} = 1.5\,10^4(R/1.5\,{\rm Mpc})$ km/s. This
is an order of magnitude larger than the ion sound speed
$c_{\rm ion}$ which is on the order of 1500 km/s,
and at which speed relativistic particles are expected to propagate through
the hot intracluster medium (Holman et al. 1979). Hence, it seems to
be more plausible that the electrons have been accelerated or, at least,
reaccelerated in situ.

{}From the discussion of the previous section we find
that, in the core region of the Coma cluster, the spectral index of
the synchrotron emission coefficient between
326 MHz and 1.38 GHz is $\alphaemn(r \lid 9\arcmin) \approx 0.4 - 0.75$
which implies an energy spectrum index of the electrons of $x \approx
1.6 - 2.5$. Considering only the integrated spectrum, Coma C is usually
classified as a steep-spectrum radio source, i.e. $\alpha > 0.75$ and hence
$x > 2.5$;
obviously, this does not apply to the halo's core region. This indicates
that there must be some very effective mechanism for particle acceleration
operating in the ICM in the core region of the cluster:
one would expect a power-law energy distribution with $x = 2.5$
if the electrons were purely originating
in a rapid leakage from radio galaxies (see, e.g., Feretti et al. 1990 for
the radio tail galaxy NGC 4869); particle acceleration in strong shocks would
produce
a power law with $x = 2$ (see e.g. Longair 1994 and references therein),
while
stochastic second order Fermi acceleration leads to an exponentially
decreasing energy spectrum (Schlickeiser 1984) which, however, may be
rather flat (effective $x < 2$) in the low-energy range
below some characteristic energy.

Recently, De Young (1992) and Tribble (1993) suggested that radio halos are
transient features associated with a major merger of two galaxy clusters
creating turbulence and shocks in the ICM: even a small conversion
efficiency should then easily accelerate high-energy particles.
This model has the great advantage to offer a natural explanation for
the rarity of the radio halo phenomenon:
after a merger event the radio emission fades
on time scales of order  $\sim 10^8$ yr due to the energy loss
of the relativistic electrons, while the time between mergers is
roughly $2-4 \times 10^9$ yr (Edge et al. 1992).
The model seems to be supported by
radio observations (Bridle \& Fomalont 1976) and X-ray observations
(Briel et al. 1991) of the merging cluster A2256 which shows that the
large diffuse radio halo is just covering the
merger region of the cluster (B\"ohringer et al. 1992; B\"ohringer
1995).
Regarding the Coma cluster,
substructures in the X-ray map as well as in the phase space distribution
of the cluster galaxies have been interpreted by White et al. (1993) and
Colless \& Dunn (1996)
as observational evidences for an ongoing merger
between the main cluster and a galaxy group dominated by NGC 4889 in
the cluster's center.
Such an ongoing merger may account for the rather
small spectral
index $\alphaemn$ in the center of the radio halo
by just enhancing the (preexisting)
turbulent velocity field in the intracluster medium (Deiss \& Just 1996),
leading to an amplification of the stochastic acceleration process (see
below).
However, it is yet unclear whether the entire extended radio halo
is caused by a single major merger as suggested by Young and Tribble and
whether the cluster's substructures observed by White et al. and Colless
\& Dunn are the remaining indications of such an event.
In contrast to what is observed in the Coma cluster, one would
expect that aging
diffuse radio halos produced in a single burst had rather irregular
and asymmetric shapes. The relaxation
time scale for a merged system of galaxies is of the order of the
galaxies' crossing time which is $\sim 10^9$ years. If
a major merger had happened only a few $10^8$ years ago,
which would be necessary considering the rather short
fading time of the halo,
one would expect a much more pronounced double-peaked phase space distribution
of the galaxies of the main cluster
than it is observed by Colless \& Dunn (1996).
Hence, even if Coma C is a transient phenomenon it can hardly  be explained
by a recent major merger of two galaxy clusters.
An additional problem is, how, in a single merger event,
electrons could be accelerated to relativistic energies out of the
thermal pool throughout the whole cluster volume.
According to standard particle-acceleration theory (e.g. Longair 1994),
a preexisting ensemble of relativistic particles is required,
in order to effectively accelerate electrons to higher energies by
shock waves.
That, in turn, suggests the requirement of the existence of some
continuously operating
reacceleration mechanism as discussed in the following.

An alternative to the cluster merger model is the
"in-situ acceleration model" proposed by Jaffe (1977). In this model,
the relativistic electrons are assumed to be continuously reaccelerated
by some mechanism operating in the (turbulent) intracluster medium.
An advantage of this model is that
it provides a natural explanation for the similarities between
diffuse radio halo
and X-ray halo of the Coma cluster, since one may expect a close link between
the
physical conditions of the relativistic particles
and that of the thermal component of the intracluster medium.
If the relativistic electrons, released from some central source,
are continuously reaccelerated a propagation speed
of $\sim 150$ km/s is sufficient to reach the
peripheral regions of the radio halo  within a Hubble time.
Giovannini et al. (1993) suggested that the orgin
of the relativistic electrons of Coma C may be the large head-tail
radio galaxy NGC 4869, orbiting at the Coma cluster center.

Deiss \& Just (1996) showed that,
due to the gravitational drag of the randomly moving galaxies,
one may expect turbulent motions $V_{\rm turb}$ of the intracluster medium
of up to 600 km/s in the core region of the Coma cluster.
This suggests that the relativistic electrons are continuously
reaccelerated by a stochastic second-order Fermi process.
According to the usual stochastic Fermi acceleration theory, the acceleration
time scale
$\tau_{\rm acc}$ is given
by $\tau_{\rm acc} = 9\kappa/<V^2>$ (e.g. Drury 1983), where $\kappa$ and $V$
denote
the spatial diffusion coefficient and the rms speed of the scattering centers
which are
presumably small-scale magnetic-field irregularities.
The diffusion coefficient is
proportional to the scattering mean free path of the particles, its value is
expected to be on the order of $10^{27} - 10^{29} {\rm cm^2/s}$. The
propagation speed
of the scatterers is basically the Alfven speed $V_{\rm Alf}$ which is on the
order of 100 km/s relative to the background medium. However, since the
magnetic field
is presumably 'frozen' in the turbulent ICM,
the scatterers' squared velocity amounts to $V^2 \approx V_{\rm turb}^2 +
V_{\rm Alf}^2$.
The correlation length $L_{\rm corr}$ of the excited stochastic velocity field
is
$\sim 20$ kpc (Deiss \& Just 1996) which is considerably larger than the
'microscopic'
scattering mean free path of the
particles. That means, in calculating the stochastic acceleration time scale,
the
usual diffusion coefficient $\kappa$ still applies. Hence, adopting
$\kappa = 10^{29} {\rm cm^2/s}$
one expects a stochastic acceleration time scale of $\tau_{\rm acc} \sim 10^7$
years at the center of the Coma cluster.
On the other hand, the more 'macroscopic'
turbulent motions generate a turbulent diffusion, hence
increasing the particles' propagation speed above the Alfven speed,
at least in the cluster's center. The excited turbulent motions scale like
$V_{\rm turb}^2 \propto n_{\rm gal}$ (Deiss \& Just 1996) where $n_{\rm gal}$
is the number density of the galaxies; hence, at a cluster
radius of $40\arcmin$ the reacceleration time scale $\tau_{\rm acc}$
is on the order of $\sim 10^9$ years. This is still fast enough
to sustain an ensemble of relativistic electrons,  although with a steeper
energy distribution than in the cluster's center. This is in accord with the
observed scale-size-vs.-frequency relation and with the steepening of the
spectral index distribution with increasing cluster
radius (see above).

In a simple leaky box model,
the steady state energy distribution of relativistic electrons,
being stochastically accelerated and losing energy via synchrotron emission and
inverse Compton scattering,
may be well approximated by (Schlickeiser et al. 1987)
\beq{N_in-situ}
  N(E) \propto E^{-\Gamma} {\rm exp}\left(-\frac{E}{E_c}\right),
\eeq
where the spectral index of the power law part and the characteristic energy
are given by
\beq{def_Gamma}
  \Gamma = \frac{3}{2}\left(1 + \frac{4\tau_{\rm acc}}{9\tau_{\rm
esc}}\right)^{1/2}
           - \frac{1}{2}
\eeq
and
\beq{def_Ec}
  E_c = 12.7 \,{\rm GeV}\left[\left(1 + \frac{2}{3}\frac{B^2}{\BCMB^2}\right)
                \frac{\tau_{\rm acc}}{{\rm 10^8 yr}}\right]^{-1},
\eeq
respectively. Since the escape time $\tau_{\rm esc}$ is on the order of the
Hubble time,
we have $\Gamma \approx 1$.
If we set $B = 6\mu$G (Feretti et
al. 1995) and $\tau_{\rm acc} = 10^7$ years in Eq. (\ref{def_Ec}) we derive a
value of
the characteristic electron energy of $E_c = 38$ GeV. The relativistic
electrons in the cluster
core, radiating at 326MHz and 1.4 GHz,
have energies of only $1.8\,(B/{\rm 6\muG})^{1/2}$
GeV and  $3.8\,(B/{\rm 6\muG})^{1/2}$ GeV respectively. This implies a value of
the
energy spectrum index of only $x \approx 1.1$ in that energy range.
In order to match
the observationally determined value of $x$ of $\sim 1.6$,
the efficiency for stochastic reacceleration may still
be even an order of magnitude smaller than inferred above.
Hence,
stochastic reacceleration of the electrons amplified by turbulent gas motion,
originating from galaxy motion inside the cluster,
appears to be sufficiently strong to account for
the rather small
radio emission index in the core region of Coma C and to sustain the
cluster-wide
distribution of the relativistic particles in that rich cluster.

In order to explain the rarity of the radio halo phenomenon, Giovannini et al.
suggested that, while the conditions for the in-situ acceleration and the
presence of a cluster-wide magnetic field (e.g. Kronberg 1994, and references
therein) are likely to be
common in rich clusters, one or more radio tail galaxies have to be present to
produce the required number of relativistic electrons.
However, there may be another constraint, namely that of a high
enough efficiency of the reacceleration mechanism which may depend
on some details in a cluster's structure; diffuse radio halos of otherwise
globally similar clusters would then have a rather dissimilar appearance. For
instance, for the core region of the Perseus cluster, Deiss \& Just (1996)
inferred
a turbulent velocity of the ICM of $\sim 200$ km/s, i.e., three times smaller
than in Coma, which implies a
ten times longer acceleration time than in the Coma cluster; if, in addition,
the magnetic field is $\sim 20 \mu$G like in other
cooling flow clusters,
i.e. about three times stronger than in Coma,
the inferred characteristic energy $E_c$ of the electron distribution
in the core region of the Perseus cluster
is two orders of magnitudes smaller than in the Coma cluster.
Obviously, such an electron distribution is not able to produce an
extended radio halo at the GHz-range, although it
may account for the observed
'minihalo' at 330 Mhz (Burns et al. 1992).
This suggests that, while a cluster-wide distribution of relativistic electrons
is likely to be common, in only a few clusters the conditions are such that
the electrons gain enough energy to produce an extended halo observable at
some hundred Megahertz. We suspect that there are far more
galaxy clusters having a diffuse radio halo than have been observed so far;
though
these halos would be observable only at rather low frequencies (well below
100 MHz), at which a systematic survey has yet to be done.
In that picture of radio halo formation, the role of a merger is that of
an amplifier of the preexisting overall stochastic reacceleration mechanism,
and not that of the prime cause of the halo formation.

The origin of the intracluster magnetic field still remains
unclear. Of course, it appears to be an attractive idea that galaxy motion
may also drive a turbulent dynamo  by which faint seed
fields could be amplified to (chaotic) microgauss fields. This suggestion
 has been
explored by Jaffe (1980), Roland (1981),  and Ruzmaikin et al. (1989). More
recently, however, De Young (1992) has shown that it is in general very
difficult
for this mechanism to produce microgauss fields on 10 kpc scales: In order to
drive a turbulent dynamo on such scales, more energetic processes, like
subcluster merging, must be envoked.
Hence, it appears that,  while reaccelerating relativistic electrons
via
galaxy motion may work, creating the intracluster magnetic field via galaxy
motion
probably does not.

\section{Summary and Conclusions}\label{sect_conclusions}

We presented new measurements of the diffuse radio emission from the
Coma cluster at 1.4 GHz using the Effelsberg single-dish telescope.
After subtracting contributions from
298 point sources we obtained
a radio map exhibiting the large scale characteristics of the
diffuse radio halo source Coma C.
{}From the large angular extent down to noise of $\sim 80\arcmin$,
one concludes that the radio halo in Coma is
a cluster-wide phenomenon which is not restricted to only the
cluster core region. This implies the presence of relativistic electrons
as well as of a magnetic field throughout a spatial volume with a
diameter of $\sim 3$ Mpc.

The integrated diffuse  flux from Coma C is $640\pm35$ mJy at 1.4 GHz. This
value fits neatly a power-law extrapolation from lower frequencies, showing
that a possible steepening of the integrated spectrum may appear only above
1.4 GHz.

We derived, under some simplifying assumptions, mutual relations between
the observationally determined spectral indices of the integrated
diffuse flux density, the surface brightness and the halo's scale size, and the
spectral
index of the synchrotron emission coefficient. These relations allow one i) to
achieve a rough estimate concerning
the consistency of the presently available data at different
frequencies and ii) to place constraints on the emissivity index distribution:
i)
We conclude that the 2.7 GHz measurements by Schlickeiser
et al. (1987) are  not consistent to our measurements. We suspect that the
value of
the integrated flux
given by these authors is much too low, which implies that the claimed strong
steepening of the spectrum above 1.4 GHz is not real.
This has implications on current theories of radio halo formation.
Hence, in order to be able to derive constraints on the physics of the
formation process of the Coma radio halo,
one needs more and improved measurements at frequencies above 1.4 GHz.
ii)
In the core region of the halo, the emissivity index
between
0.3 GHz and 1.4 GHz appears to be in the range 0.4 - 0.75
which implies an energy spectrum index of the electrons of $x \approx
1.6 - 2.5$. From that we conclude
that there must be some very effective mechanism for particle acceleration
operating in the intracluster medium.

We argued that the observed large extent of Coma C, its regular shape as
well as  its clear
similarity with the X-ray halo support an in-situ acceleration model for
radio halo formation,
 since one may expect a close link between the
physical conditions of the relativistic particles
and that of the thermal component of the intracluster medium.
We showed that in the Coma cluster
stochastic reacceleration of the electrons, amplified by turbulent gas motion
originating from galaxy motion inside the cluster (Deiss \& Just 1996),
appears to be sufficiently strong to account for
the inferred rather small
radio emission index in the core region of Coma C and to sustain a cluster-wide
distribution of relativistic particles in that rich cluster.

We suggested that the rarity of the radio halo phenomenon has its origin in
that
the efficiency of the stochastic reacceleration mechanism may depend
on some details of the clusters' structure: diffuse radio halos
of otherwise globally similar clusters may have a rather dissimilar appearance.
That means, while a cluster-wide distribution of relativistic electrons
is likely to be common in rich clusters, in only a few of them the conditions
are such that
the electrons gain enough energy to produce an extended halo observable at
some hundred Megahertz. However, we suspect that at low
frequencies (well below 100 MHz) diffuse radio halos of galaxy clusters
are much more common
than it is suggested by the presently available data.

\acknowledgements
{We like to thank Dr. Tom Landecker for making the list of radio sources
from the Coma cluster of galaxies available to us in digital form.}

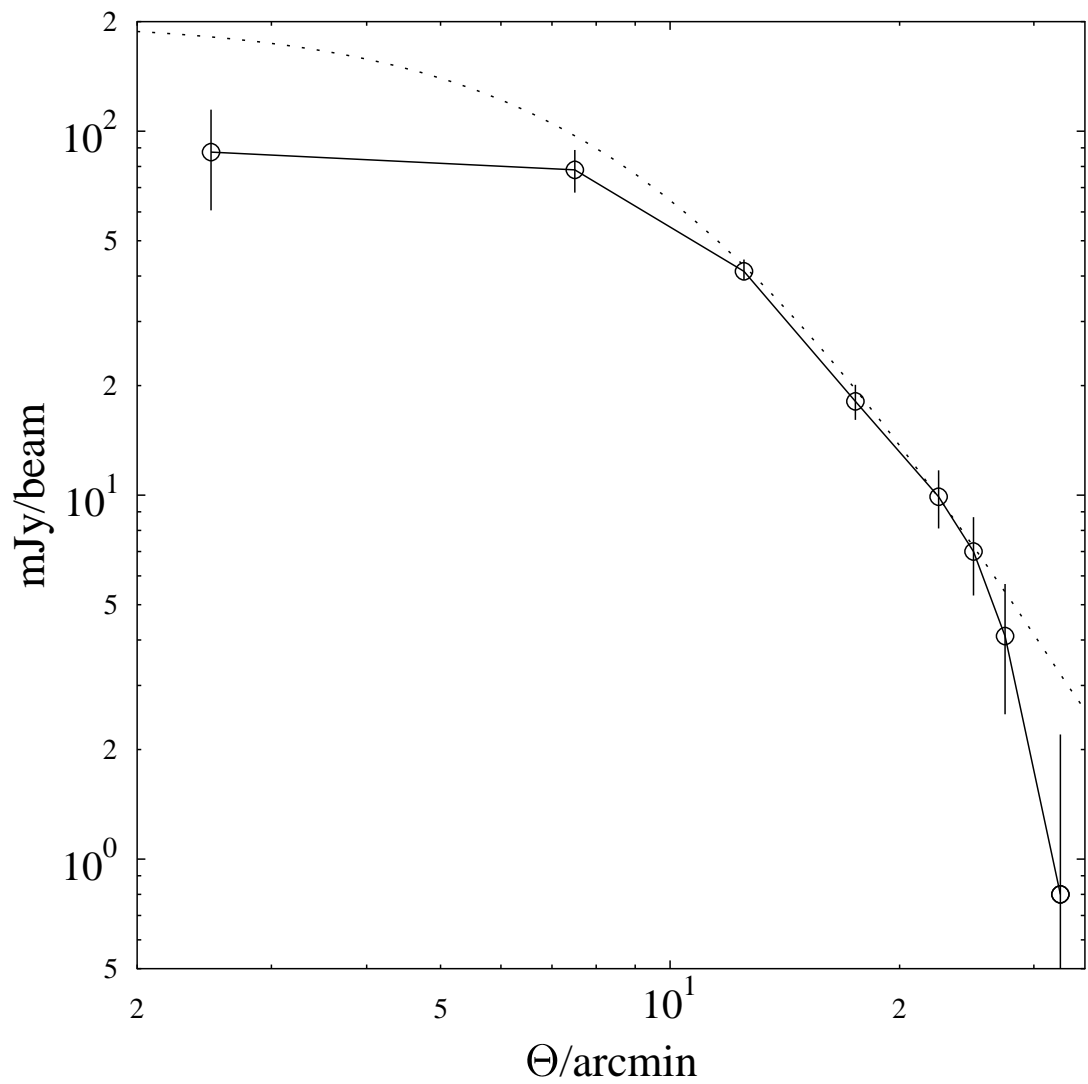
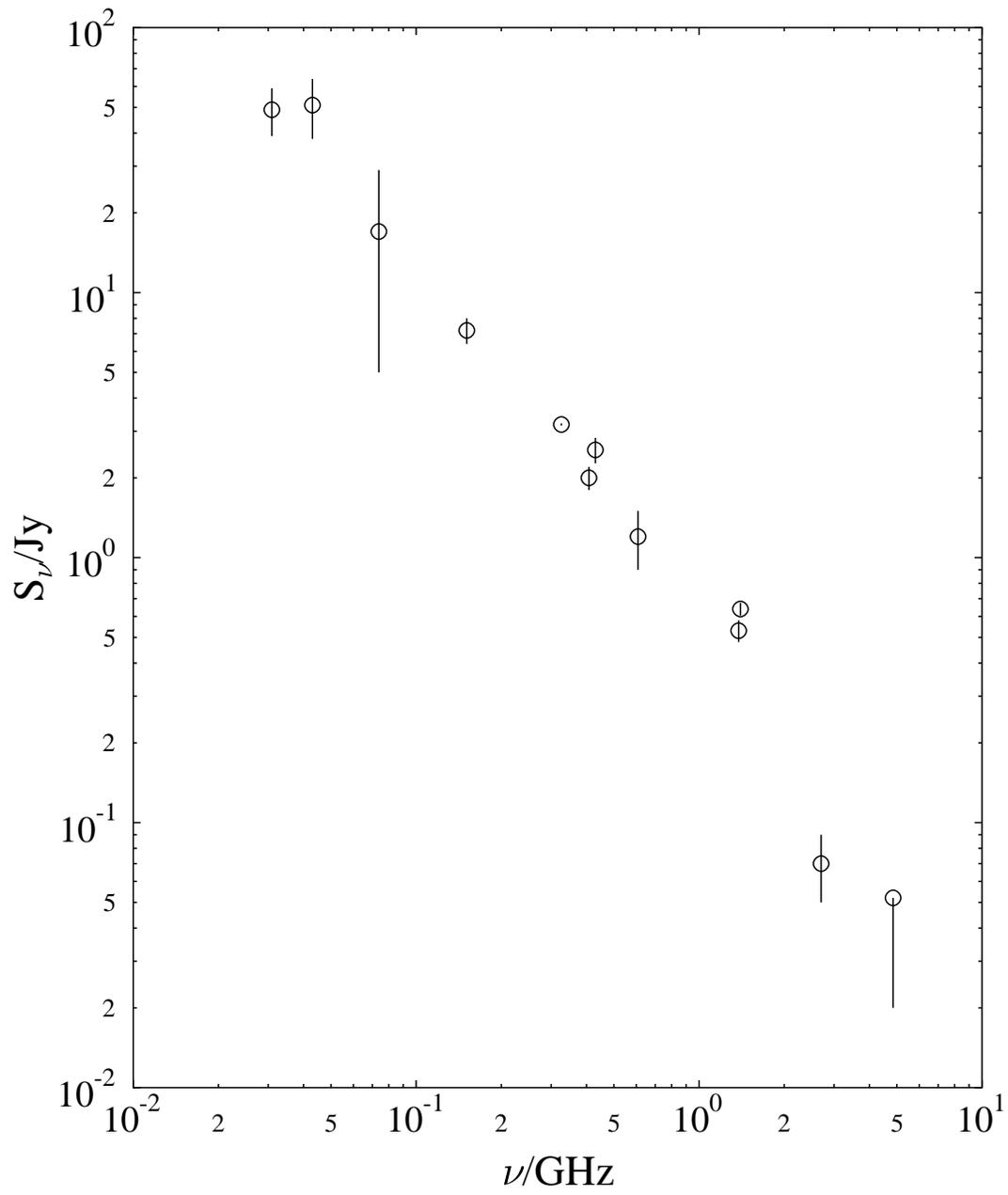
\end{document}